\documentclass[useAMS,usenatbib,a4paper,referee]{mn2e}
\usepackage{graphicx,times}
\usepackage{amsmath}

\title[Angular Size Test]
  {Cosmological Tests using the Angular Size of Galaxy Clusters}
  \author[Wei, Wu \& Melia]
    {Jun-Jie Wei$^{1,2}$\thanks{Email:jjwei@pmo.ac.cn}, Xue-Feng Wu$^{1,3,4}$\thanks{Email:xfwu@pmo.ac.cn}, and Fulvio Melia$^{1,5}$\thanks{John Woodruff Simpson Fellow. Email: fmelia@email.arizona.edu} \\
  $^1$Purple Mountain Observatory, Chinese Academy of Sciences, Nanjing 210008, China\\
  $^2$University of Chinese Academy of Sciences, Beijing 100049, China\\
  $^3$Chinese Center for Antarctic Astronomy, Nanjing 210008, China\\
  $^4$Joint Center for Particle, Nuclear Physics and Cosmology, Nanjing University-Purple Mountain Observatory, Nanjing 210008, China\\
  $^5$Department of Physics, The Applied Math Program, and Department of Astronomy, The University of Arizona, AZ 85721, USA}

\begin{document}


%

\maketitle

\begin{abstract}
We use measurements of the galaxy-cluster angular size versus redshift
to test and compare the standard model ($\Lambda$CDM) and the
$R_{\rm h}=ct$ Universe. We show that the latter fits the data with
a reduced $\chi^2_{\rm dof}=0.786$ for a Hubble constant $H_{0}=
72.6_{-3.4}^{+3.8}$ km $\rm s^{-1}$ $\rm Mpc^{-1}$, and $H_{0}$ is
the sole parameter in this model. By comparison, the optimal
flat $\Lambda$CDM model, with two free parameters (including
$\Omega_{\rm m}=0.50$ and $H_{0}=73.9_{-9.5}^{+10.6}$ km $\rm s^{-1}$
$\rm Mpc^{-1}$), fits the angular-size data with a reduced
$\chi^2_{\rm dof}=0.806$. On the basis of their $\chi^2_{\rm dof}$
values alone, both models appear to account for the data very well
in spite of the fact that the $R_{\rm h}=ct$ Universe expands
at a constant rate, while $\Lambda$CDM does not. However, because 
of the different number of free parameters in these models, selection 
tools, such as the Bayes Information Criterion, favour $R_{\rm h}=ct$ over
$\Lambda$CDM with a likelihood of $\sim 86\%$ versus $\sim 14\%$. 
These results impact the question of galaxy growth at large redshifts. 
Previous work suggested an inconsistency with the underlying 
cosmological model unless elliptical and disk galaxies grew in size 
by a surprisingly large factor $\sim 6$ from $z\sim 3$ to $0$. The fact 
that both $\Lambda$CDM and $R_{\rm h}=ct$ fit the cluster-size measurements
quite well casts some doubt on the suggestion that the unexpected
result with individual galaxies may be due to the use of an incorrect
expansion scenario, rather than astrophysical causes, such as mergers
and/or selection effects.
\end{abstract}

\begin{keywords}
Cosmology: theory, observations, large-scale structure---galaxies: clusters: general---galaxies:evolution
\end{keywords}
\section{Introduction}
The use of standard rods---distance scales with no evolution in linear
size over the lifetime of the Universe---to carry out geometric tests of
cosmological models was first proposed by Hoyle (1959), but the application
of this idea to real data took a long time, specifically because of the
difficulty in finding suitable objects or structures for this purpose.
The earliest tests of cosmological models using the observed dependence
of the angular size of galaxies or kpc-scale radio sources could not easily
define a true metric rod and were subject to unknown evolutionary
effects (e.g., Sandage 1988). There was uncertainty about whether the
observed size-redshift relation in radio galaxies was an indication of an
actual evolution in size (Kapahi 1987; Barthel \& Miley 1988; Neeser et al.
1995), or whether it was due to selection effects (Singal 1998; Nilsson
et al. 1993).  This was partially addressed in a subsequent study using
an enlarged sample of double-lobed quasars at $z>0.3$ (Buchalter et al.
1998), which showed no change in apparent angular size within the
range $1.0< z < 2.7$, consistent with standard cosmology without
significant evolution.

Separate investigations specifically to study the cosmic deceleration
(or acceleration) were carried out by Gurvits (1993, 1994) (using VLBI
visibility data obtained at 13 cm by Preston et al. 1985). This analysis
also provided estimates of the dependence of the apparent angular
size of compact sources on their luminosity and rest-frame frequency.
Similar studies were also carried out by Kellermann (1993) and
Wilkinson et al. (1998). Within a few years, a much enlarged sample
of 330 compact radio sources distributed over a broad range of redshifts
$0.011< z <4.72$ started to demonstrate that the angular size-redshift
relation for compact radio sources is consistent with the predictions
of standard Friedmann-Robertson-Walker models without the need
to consider evolutionary or selection effects (Gurvits et al. 1999).

Other groups carried out their own cosmological tests using powerful
radio-lobed radio galaxies, presumed to be reasonable standard
yardsticks to determine global cosmological parameters (Daly 1994, 1995).
The method was applied and discussed by Guerra \& Daly (1996, 1998),
Guerra (1997), and Daly, Guerra \& Wan (1998, 2000), who reported that
the data at that time strongly favoured a low (matter) density Universe.

Vishwakarma (2001) and Lima \& Alcaniz (2002) used the Gurvits et al.
(1999) compact radio source angular size versus redshift data to set
constraints on cosmological parameters. This analysis was extended
by Chen \& Ratra (2003), who used these data to place constraints
on cosmological model parameters for a variety of cosmological constant
scenarios. And a more focused study using FRIIb radio galaxy
redshift-angular size data to derive constraints on the parameters
of a spatially-flat cosmology with a dark-energy scalar field was
carried out by Podariu et al. (2003).

More recently, in one of the better known studies using this method,
based on the average linear size of galaxies with the same luminosity
(see, e.g., McIntosh et al. 2005; Barden et al. 2005; Trujillo et al. 2006;
L\'opez-Corredoira 2010), the measurements do not appear to be consistent
with an expanding cosmology, unless galaxies have grown in size
by a surprisingly large factor six from redshift $z=3.2$ to $z=0$.

Perhaps the cosmology itself is wrong or, more simply, there is still
some ambiguity concerning the use of galactic size as a standard rod.
Indeed, if another more reliable scale could be found, and shown to
be consistent with, say, the standard model ($\Lambda$CDM), the
contrast between this result, and the disparity emerging through
the use of galaxies, could be useful in affirming the need for
stronger evolution in galactic growth than is predicted by current
theory.

Our focus in this paper will be galaxy clusters, which can also
be used as standard rulers under appropriate
conditions. These are the largest gravitationally collapsed structures
in the Universe, with a hot diffuse plasma ($T_{e}\sim10^{7}-10^{8}$ K)
filling the intergalactic medium. Their angular size versus redshift
can be measured using a combination of the Sunyaev-Zel'dovich effect
(SZE) and X-ray surface brightness observations. The Sunyaev-Zel'dovich
effect is the result of high-energy electrons distorting the cosmic
microwave background radiation (CMB) through inverse Compton scattering,
during which low-energy CMB photons (on average) receive an energy boost
from the high-energy electrons in the cluster (Sunyaev \& Zel'dovich
1970, 1972). The same hot gas emits X-rays primarily via thermal
bremsstrahlung. While the SZE is a function of the integrated pressure,
$\Delta T \propto \int n_{e}T_{e}\;dl$ (in terms of the electron number
density $n_{e}$ and temperature $T_{e}$) along the line-of-sight,
the X-ray emission scales as $S_{X}\propto \int n_{e}^{2}
\Lambda_{ee}\;dl$ (in terms of the cooling function $\Lambda_{ee}$).
This different dependence on density, along with a suitable model for
the cluster gas, enables a direct distance determination to the galaxy
cluster. This method is independent of the extragalactic distance ladder
and provides distances to high-redshift galaxy clusters.

SZE/X-ray distances have been previously used to constrain some
cosmological parameters and to test the distance duality relationship
of metric gravity models (see, e.g., De Bernardis et al. 2006;
Lima et al. 2010; Cao \& Liang 2011; Holanda et al. 2012; Chen \& Ratra
2012; Liang et al. 2013; Chen et al. 2013). In this paper,
we will use these data, not to further examine the distance duality
relation per se but, rather, to address two related issues. First,
we will use the newer and larger sample of galaxy-cluster angular size
versus redshift measurements from Bonamante et al. (2006) to constrain
cosmological models. In particular, we wish to see if the aforementioned
tension between galaxy growth and the conventional expansion scenario is
supported by the cluster data, or whether the latter confirm the basic
theoretical predictions, thus reinforcing the need for a stronger galactic
evolution at high redshifts. Second, we wish to use this relatively new
probe of the Universe's expansion to directly test the $R_{\rm h}=ct$
Universe (Melia 2007; Melia \& Shevchuk 2012) against the data and to
see how its predictions compare with those of $\Lambda$CDM.

The outline of this paper is as follows. In \S~2, we will briefly
summarize the galaxy-cluster angular-size sample at our disposal.
We will present theoretical fits to the data in \S~3, and constrain
the cosmological parameters---both in the context of $\Lambda$CDM
and the $R_{\rm h}=ct$ Universe---in \S~4. We will end with a
discussion and conclusion in \S~5.

\section{The Cluster Angular-Size Sample}
In addition to the luminosity distance, $D_{L}$, which is necessary
for measurements involving standard candles such as Type Ia SNe,
the angular-diameter distance (ADD), $D_A$, is also used in astronomy
for objects whose diameter (i.e., the standard ruler) is known.
The luminosity distance and ADD can be measured independently using
different celestial objects, but they are related via Etherington's
reciprocity relation:
\begin{equation}
\frac{D_{L}}{D_{A}}(1+z)^{-2}=1.
\label{distance}
\end{equation}
This relation, sometimes referred as the distance-duality (DD) relation,
is completely general and is valid for all cosmological models based on
Riemannian geometry. That is, its validity is independent of Einstein's
field equations for gravity and the nature of the matter-energy content
of the universe. It requires only that the source and observer be connected
via null geodesics in a Riemannian spacetime and that the number of
photons be conserved.

X-ray observations of the intra-cluster medium, combined with radio
observations of the galaxy cluster's Sunyaev-Zel'dovich effect,
allow an estimate of the ADD to be made. Recently, Bonamante
et al. (2006) determined the distance to 38 clusters of galaxies
in the redshift range $0.14\leq z \leq 0.89$ using X-ray data from
\emph{Chandra} and SZE data from the Owens Valley Radio Observatory
and the Berkeley-Illinois-Maryland Association interferometric arrays.
The data shown in Table~1 are reproduced from the compilation
of Bonamante et al. (2006).

\begin{table}
\centering
\centerline{{\bf Table 1. } Angular Diameter Distance of Galaxy Clusters}
\label{1}
\begin{tabular}{llr}
\\ \hline
Cluster\qquad\qquad\qquad\qquad\qquad&$\;\;\,z$&\qquad$D_{A}(\rm Mpc)$\\
\hline
Abell	1413	&	0.142	&$	780	^{+	180	}_{-	130	}$\\
Abell	2204	&	0.152	&$	610	^{+	60	}_{-	70	}\;\,$\\
Abell	2259	&	0.164	&$	580	^{+	290	}_{-	250	}$\\
Abell	586	&	0.171	&$	520	^{+	150	}_{-	120	}$\\
Abell	1914	&	0.171	&$	440	^{+	40	}_{-	50	}$\;\,\\
Abell	2218	&	0.176	&$	660	^{+	140	}_{-	110	}$\\
Abell	665	&	0.182	&$	660	^{+	90	}_{-	100	}$\\
Abell	1689	&	0.183	&$	650	^{+	90	}_{-	90	}$\;\,\\
Abell	2163	&	0.202	&$	520	^{+	40	}_{-	50	}$\;\,\\
Abell	773	&	0.217	&$	980	^{+	170	}_{-	140	}$\\
Abell	2261	&	0.224	&$	730	^{+	200	}_{-	130	}$\\
Abell	2111	&	0.229	&$	640	^{+	200	}_{-	170	}$\\
Abell	267	&	0.23	&$	600	^{+	110	}_{-	90	}$\\
RX	J2129.7+0005	&	0.235	&$	460	^{+	110	}_{-	80	}$\\
Abell	1835	&	0.252	&$	1070	^{+	20	}_{-	80	}$\;\,\\
Abell	68	&	0.255	&$	630	^{+	160	}_{-	190	}$\\
Abell	697	&	0.282	&$	880	^{+	300	}_{-	230	}$\\
Abell	611	&	0.288	&$	780	^{+	180	}_{-	180	}$\\
ZW	3146	&	0.291	&$	830	^{+	20	}_{-	20	}$\;\,\\
Abell	1995	&	0.322	&$	1190	^{+	150	}_{-	140	}$\\
MS	1358.4+6245	&	0.327	&$	1130	^{+	90	}_{-	100	}$\\
Abell	370	&	0.375	&$	1080	^{+	190	}_{-	200	}$\\
MACS	J2228.5+2036	&	0.412	&$	1220	^{+	240	}_{-	230	}$\\
RX	J1347.5-1145	&	0.451	&$	960	^{+	60	}_{-	80	}$\;\,\\
MACS	J2214.9-1359	&	0.483	&$	1440	^{+	270	}_{-	230	}$\\
MACS	J1311.0-0310	&	0.49	&$	1380	^{+	470	}_{-	370	}$\\
CL	0016+1609	&	0.541	&$	1380	^{+	220	}_{-	220	}$\\
MACS	J1149.5+2223	&	0.544	&$	800	^{+	190	}_{-	160	}$\\
MACS	J1423.8+2404	&	0.545	&$	1490	^{+	60	}_{-	30	}$\;\,\\
MS	0451.6-0305	&	0.55	&$	1420	^{+	260	}_{-	230	}$\\
MACS	J2129.4-0741	&	0.57	&$	1330	^{+	370	}_{-	280	}$\\
MS	2053.7-0449	&	0.583	&$	2480	^{+	410	}_{-	440	}$\\
MACS	J0647.7+7015	&	0.584	&$	770	^{+	210	}_{-	180	}$\\
MACS	J0744.8+3927	&	0.686	&$	1680	^{+	480	}_{-	380	}$\\
MS	1137.5+6625	&	0.784	&$	2850	^{+	520	}_{-	630	}$\\
RX	J1716.4+6708	&	0.813	&$	1040	^{+	510	}_{-	430	}$\\
MS	1054.5-0321	&	0.826	&$	1330	^{+	280	}_{-	260	}$\\
CL	J1226.9+3332	&	0.89	&$	1080	^{+	420	}_{-	280	}$\\
\hline
\end{tabular}
\end{table}

\section{Theoretical Fits}
The theoretical angular diameter distance $D_{A}$ is a function of the cluster's
redshift \emph{z}, and is different for different cosmological models. Both
$\Lambda$CDM and $R_{\rm h}=ct$ are FRW cosmologies,
but the latter includes the additional constraint $p=-\rho/3$ on the overall
equation of state. The densities are often written in terms of today's
critical density, $\rho_c\equiv 3c^2 H_0^2/8\pi G$, represented as
$\Omega_{\rm m}\equiv\rho_{\rm m}/\rho_c$, $\Omega_{\rm r}\equiv\rho_{\rm r}/
\rho_c$, and $\Omega_{\rm de}\equiv \rho_{\rm de}/\rho_c$. $H_0$ is the Hubble
constant, and $\Omega_{\rm de}$ is simply $\Omega_\Lambda$ when dark energy is
assumed to be a cosmological constant.

\subsection{$\Lambda$CDM}
In a flat $\Lambda$CDM Universe with zero
spatial curvature, the total scaled energy density is $\Omega\equiv
\Omega_{\rm m}+\Omega_{\rm r}+\Omega_{\rm de}=1$.
When dark energy is included with an unknown equation of state,
$p_{\rm de}=w_{\rm de}\rho_{\rm de}$, the most general form of the
angular diameter distance is given by the expression
\begin{eqnarray}
\qquad D_{A}^{\Lambda {\rm CDM}}(z)&=&{1\over H_{0}}{c\over\mid\Omega_{k}\mid^{1/2}(1+z)}
\; {\rm sinn}\Bigg\{\mid\Omega_{k}\mid^{1/2}\nonumber \\
&\null&\qquad\left.\times\int_{0}^{z}{dz\over
\sqrt{\Omega_{\rm m}(1+z)^{3}+\Omega_{k}(1+z)^{2}+\Omega_{\rm de}
(1+z)^{3(1+w_{\rm de})}}}\right\}\;,
\end{eqnarray}
where $c$ is the speed of light. In this equation,
$\Omega_{k}=1-\Omega_{\rm m}-\Omega_{\rm de}$ represents the
spatial curvature of the Universe---appearing as a term proportional
to the spatial curvature constant $k$ in the Friedmann equation. In
addition, sinn is $\sinh$ when $\Omega_{k}>0$ and $\sin$ when $\Omega_{k}<0$.
For a flat Universe with $\Omega_{k}=0$, the right-hand expression simplifies to
the form $(c/H_{0})(1+z)^{-1}$ times the integral.

To cover a reasonable representation of the parameter space in
$\Lambda$CDM, we consider the following two models for dark energy,
and one without:

1. $\Lambda$CDM---the ``standard" model, with a cosmological constant
in a flat universe. The dark-energy equation of state parameter, $w_{\rm de}$,
is exactly $-1$, and $\Omega_{\rm de}\equiv\Omega_{\Lambda}=1-\Omega_{\rm m}$
(since radiation is negligible at low redshifts).

2. $w$CDM---a flat universe with a constant dark-energy equation of state,
but with a $w_{\rm de}$ that is not necessarily equal to $-1$. Here
$\Omega_{k}=0$, so the free parameters may be chosen from the
following: $H_0$, $\Omega_{\rm m}$ and $w_{\rm de}$.

3. For comparison, we also consider a model with $\Omega_{\rm m}=1.0$, i.e.,
the Einstein-de Sitter cosmology.

\subsection{The $R_{\rm h}=ct$ Universe}
In the $R_{\rm h}=ct$
Universe, the angular diameter distance is given by the much simpler expression
\begin{equation}
D_{A}^{R_{\rm h}=ct}(z)=\frac{c}{H_{0}}\frac{\ln(1+z)}{1+z}\;.
\end{equation}
The factor $c/H_0$ is in fact the gravitational horizon $R_{\rm h}(t_0)$ at the
present time, so we may also write the angular diameter distance as
\begin{equation}
D_{A}^{R_{\rm h}=ct}(z)=R_{\rm h}(t_0)\frac{\ln(1+z)}{1+z}\;.
\end{equation}
A detailed account of the differences between $\Lambda$CDM and
$R_{\rm h}=ct$ is provided in Melia \& Shevchuk (2012), Melia \& Maier
(2013), and Wei et al. (2013). A more pedagogical description may also
be found in Melia (2012). An important distinction between these
two models is that whereas the $R_{\rm h}=ct$ Universe expands at
a constant rate, $\Lambda$CDM predicts an early phase of deceleration,
followed by a current acceleration. Therefore, an examination of the
cluster angular size data, spanning the redshift range ($\sim 0<z<0.9$)
within which the transition from deceleration to acceleration is thought 
to have occurred, could in principle help to distinguish between these
two cosmologies. 

Briefly, the $R_{\rm h}=ct$ Universe is a Friedmann-Robertson-Walker (FRW)
cosmology that has much in common with $\Lambda$CDM, but includes an additional
ingredient motivated by several theoretical and observational arguments
(Melia 2007; Melia \& Abdelqader 2009; Melia \& Shevchuk 2012; Melia 2013a). Like
$\Lambda$CDM, it adopts an equation of state $p=w\rho$, with $p=p_{\rm m}+
p_{\rm r}+p_{\rm de}$ and $\rho=\rho_{\rm m}+\rho_{\rm r}+\rho_{\rm de}$, but
goes one step further by specifying that $w=(\rho_{\rm r}/3+w_{\rm de}
\rho_{\rm de})/\rho=-1/3$ at all times. Here, $p$ is the pressure and
$\rho$ is the energy density, and subscripts r, m, and de refer to
radiation, matter, and dark energy, respectively. One might come away with the
impression that this equation of state cannot be consistent with that
(i.e., $w=[\rho_{\rm r}/3-\rho_\Lambda]/\rho$) in the standard model. But
in fact nature is telling us that if we ignore the constraint $w=-1/3$ and
instead proceed to optimize the parameters in $\Lambda$CDM by fitting
the data, the resultant value of $w$ averaged over a Hubble time is
actually $-1/3$ within the measurement errors (Melia 2007; Melia \&
Shevchuk 2012). In other words, though $w=(\rho_{\rm r}/3-\rho_\Lambda)/\rho$
in $\Lambda$CDM cannot be equal to $-1/3$ from one moment to the next, its
value averaged over the age of the Universe is equal to what it would have
been in $R_{\rm h}=ct$.

In terms of the expansion dynamics, $\Lambda$CDM must guess the constituents
of the Universe and their individual equations of state, and then predict
the expansion rate as a function of time. In contrast, $R_{\rm h}=ct$
acknowledges the fact that no matter what these constituents are, the
total energy density in the Universe gives rise to a gravitational
horizon coincident with the better known Hubble radius. But because
this radius is therefore a proper distance, the application of Weyl's
postulate forces it to always equal $ct$. Thus, on every time slice,
the energy density $\rho$ must partition itself among its various
constituents ($\rho_{\rm m}$, $\rho_{\rm r}$ and $\rho_{\rm de}$) in
such a way as to always adhere to this constraint, which also
guarantees that the expansion rate be constant in time.

\subsection{Optimization of the Parameters}
For each model, the best-fit is obtained by minimizing the function
\begin{equation}
\begin{split}
\chi^{2}_{\rm ADD}=\sum_{i=1}^{38}\frac{[D_{A}^{\rm th}(z_{i},\xi)-D_{A}^{\rm obs}(z_{i})]^{2}}
{\sigma^{2}_{\rm tot, \emph{i}}}\;,
\end{split}
\end{equation}
where $\xi$ stands for all the cosmological parameters that define the fitted
model, $z_i$ is the redshift of the observed galaxy cluster, $D_{A}^{\rm th}$
is the predicted value of the ADD in the cosmological model under consideration,
and $D_{A}^{\rm obs}$ is the measured value. There are three sources of uncertainty
in the measurement of $D_{A}$: the cluster modeling error $\sigma_{\rm mod}$,
the statistical error $\sigma_{\rm stat}$, and the systematic error
$\sigma_{\rm sys}$. The modeling errors are shown in Table~1 and the
statistical and systematic errors are presented in Table~3 of Bonamante
et al. (2006). In our analysis, we combine these errors in quadrature. Thus,
the total uncertainty $\sigma_{\rm tot}$ is given by the expression
$\sigma_{\rm tot}^{2}=\sigma_{\rm mod}^{2}+\sigma_{\rm stat}^{2}
+\sigma_{\rm sys}^{2}$.

\begin{figure}
\vskip 0.2in
\centerline{\includegraphics[angle=0,scale=0.8]{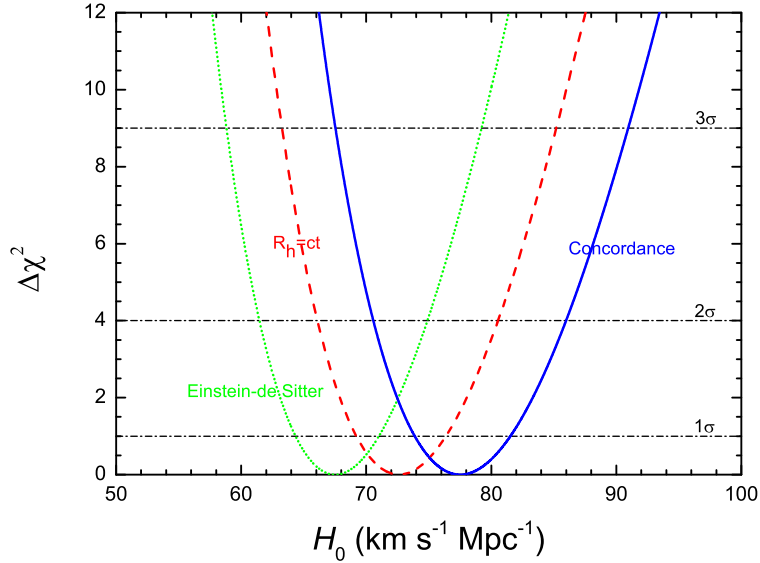}}
\caption{Constraints on the Hubble parameter, $H_{0}$, for the $R_{\rm h}=ct$
(\emph{dashed}), Einstein-de Sitter (\emph{dotted}), and concordance (\emph{solid})
models, respectively.}\label{Rh}
\end{figure}

\section{Results}

Though the number of free parameters characterizing $\Lambda$CDM can
be as large as 6 or 7, depending on the application, here we take the
minimalist approach and use the four most essential ones: the Hubble
constant $H_{0}$, the matter energy density $\Omega_{\rm m}$, the
dark-energy density $\Omega_{\rm de}$, and the dark energy equation
of state parameter $w_{\rm de}$. By comparison, the $R_{\rm h}=ct$
Universe has only one free parameter---the Hubble constant $H_{0}$.

\begin{figure}
\vskip 0.2in
\centerline{\includegraphics[angle=0,scale=0.8]{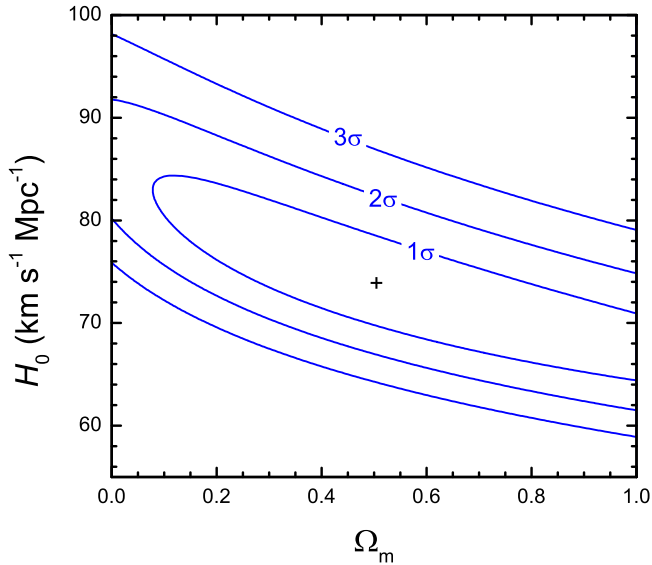}}
\caption{$1\sigma-3\sigma$ constraint contours for the flat $\Lambda$CDM
model, using the Bonamante et al. (2006) ADD data. The cross indicates
the best-fit pair ($\Omega_{\rm m}$, $H_{0}$)=(0.50, 73.93 km $\rm
s^{-1}$ $\rm Mpc^{-1}$).}\label{flat}
\end{figure}

In this section, we optimize the fit for several dark-energy models
(which we call $\Lambda$CDM and $w$CDM), one without dark energy (the
Einstein-de Sitter Universe), and for the $R_{\rm h}=ct$ Universe.
The outcome for each model is described and discussed in subsequent
subsections.

\subsection{$\Lambda$CDM}
In $\Lambda$CDM, the dark-energy equation of state parameter, $w_{\rm de}\equiv
w_\Lambda$, is exactly $-1$. Assuming a flat universe, $\Omega_{\Lambda}=1-
\Omega_{\rm m}$, there are only two free parameters: $\Omega_{\rm m}$ and
$H_{0}$. Type Ia SN measurements (see, e.g., Perlmutter et al.  1998,
1999; Schmidt et al. 1998; Riess et al. 1998; Garnavich et al. 1998),
CMB anisotropy data (see, e.g., Ratra et al. 1999; Podariu et al.
2001; Spergel et al. 2003; Komatsu et al. 2009, 2011; Hinshaw et al. 2013), and baryon
acoustic oscillation (BAO) peak length scale estimates (see, e.g.,
Percival et al. 2007; Gazta{\~n}aga et al. 2009; Samushia \& Ratra 2009),
strongly suggest that we live in a spatially flat, dark energy-dominated unverse
with concordance parameter values $\Omega_{\rm m}\approx0.3$ and $\Omega_{\rm de}
\approx0.7$. We will first adopt this concordance model, though to improve the
fit we keep $H_{0}$ as a free parameter. The 38 cluster distances are fit with
this theoretical $D_{A}(z)$ function and the constraints on $H_{0}$ are shown
in Figure~1 (solid curve). For this fit, we obtain $H_{0}=77.5_{-3.6}^{+4.0}$
km $\rm s^{-1}$ $\rm Mpc^{-1}$ ($68\%$ confidence interval). The $\chi^{2}$
per degree of freedom for the concordance model with an optimized Hubble
constant is $\chi_{\rm dof}^{2}=29.21/37=0.789$, remembering that all of its
parameters, save for $H_0$, are assumed to have prior values.

If we relax the priors, and allow both $\Omega_{\rm m}$ and $H_{0}$
to be free parameters, we obtain best-fit values $(\Omega_{\rm m}$, $H_{0}$)
=(0.50, $73.9_{-9.5}^{+10.6}$ km $\rm s^{-1}$ $\rm Mpc^{-1}$).
Figure~2 shows the $1\sigma-3\sigma$ constraint contours of the probability
in the ($\Omega_{\rm m}$, $H_{0}$) plane. These contours show that at the
$1\sigma$ level, $64.4<H_{0}<84.5$ km $\rm s^{-1}$ $\rm Mpc^{-1}$, but that $\Omega_{\rm m}$ is poorly
constrained; only a lower limit of $\sim0.08$ can be set at this confidence
level. The cross indicates the best-fit pair ($\Omega_{\rm m}$, $H_{0}$)
=(0.50, 73.9 km $\rm s^{-1}$ $\rm Mpc^{-1}$). We find that the $\chi^{2}$
per degree of freedom is $\chi_{\rm dof}^{2}=29.01/36=0.806$.

\subsection{$w$CDM and Einstein-de Sitter}
For the $w$CDM model, $w_{\rm de}$ is
constant but possibly different from $-1$. For a flat universe
($\Omega_{k}=0$), there are therefore possibly three free parameters:
$\Omega_{\rm m}$, $w_{\rm de}$, and $H_{0}$, though to keep the
discussion as simple as possible, we will here treat
$\Omega_{\rm m}$ as fixed at the value $0.3$, and allow
$H_0$ and $w_{\rm de}$ to vary.

\begin{figure}
\vskip 0.2in
\centerline{\includegraphics[angle=0,scale=0.8]{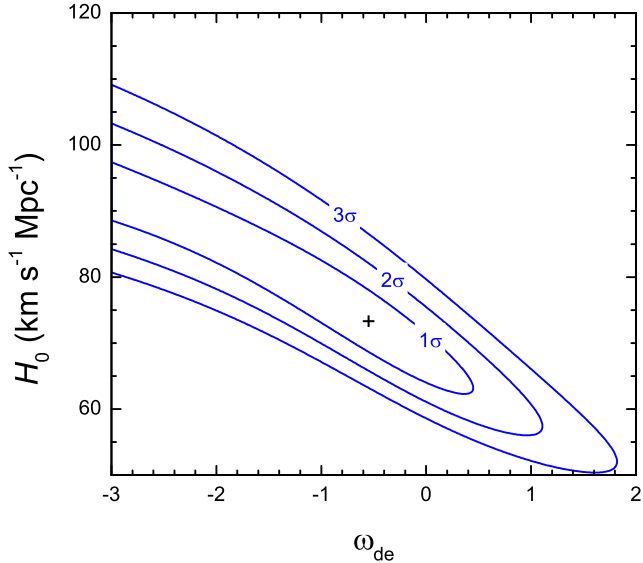}}
\caption{$1\sigma-3\sigma$ constraint contours for the $w$CDM
model ($\Omega_{\rm m}$ is fixed to be 0.3), using the Bonamante et al. (2006) ADD data. The cross
indicates the best-fit pair ($w_{\rm de}$, $H_{0}$)=
($-0.55$, 73.1 km $\rm s^{-1}$ $\rm Mpc^{-1}$).}\label{wCDM}
\end{figure}

Figure~3 shows the constraints using the Bonamante et al. (2006)
ADD data with the dark-energy model. These contours show that at the
$1\sigma$ level, $62.3<H_{0}<97.3$ km $\rm s^{-1}$ $\rm Mpc^{-1}$,
 but that $w_{\rm de}$ is poorly
constrained; only an upper limit of $\sim0.47$ can be set at this confidence
level. The cross indicates the best-fit pair ($w_{\rm de}$, $H_{0}$)
=($-0.55$, 73.1 km $\rm s^{-1}$ $\rm Mpc^{-1}$) with
$\chi_{\rm dof}^{2}=29.04/36=0.807$. With the current level of precision,
we see that the cluster ADD data can be fit very well with this
variation of $\Lambda$CDM, but the optimized parameter $w_{\rm de}$ is only
marginally consistent with that of the concordance model. We emphasize, however,
that the scatter seen in the best-fit diagram (Figure~4) is still
significant. Future measurements of the ADD may provide much tighter
constraints on the cosmological parameters.

For the Einstein-de Sitter (E-deS) model with $\Omega_{\rm m}=1$, the
only free parameter is $H_{0}$. As shown in Figure~1 (dotted curve),
the best fit corresponds to $H_{0}=67.5_{-3.2}^{+3.5}$ km $\rm s^{-1}$
$\rm Mpc^{-1}$ ($68\%$ confidence interval). With $38-1=37$ degrees of
freedom, the reduced $\chi^{2}$ for the Einstein-de Sitter model with
an optimized Hubble constant is $\chi_{\rm dof}^{2}=29.49/37=0.797$.

\begin{figure}
\vskip 0.2in
\centerline{\includegraphics[angle=0,scale=1.0]{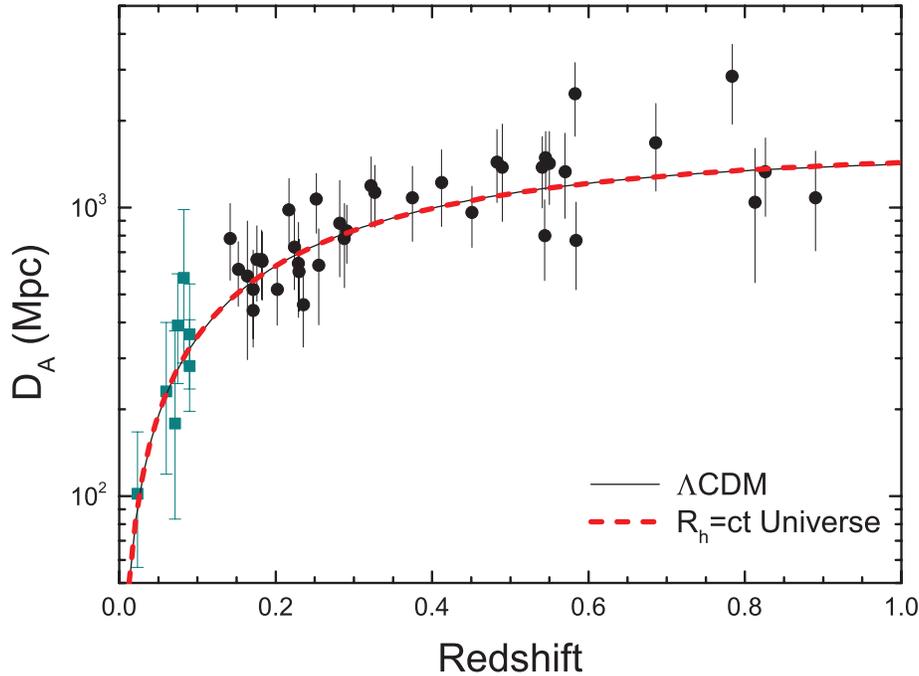}}
\caption{ADD of the 38 clusters (\emph{points}), with error bars,
and the best fit theoretical curves: (\emph{dashed}) the
$R_{\rm h}=ct$ Universe, with its sole optimized parameter
$H_{0}=72.6_{-3.4}^{+3.8}$ km $\rm s^{-1}$ $\rm Mpc^{-1}$,
and (\emph{solid}) the standard, flat $\Lambda$CDM cosmology,
with $\Omega_{\rm m}=0.50$ and $H_{0}=73.9_{-9.5}^{+10.6}$
km $\rm s^{-1}$ $\rm Mpc^{-1}$. The reduced $\chi_{\rm dof}^{2}$
(with 37 degrees of freedom) for the $R_{\rm h}=ct$ fit is
0.786. The corresponding value for the optimal $\Lambda$CDM model
(with 36 degrees of freedom) is $\chi_{\rm dof}^{2}=0.806$.
The squares are from the low-redshift sample of Mason et al.
(2001); these are not included in the fits, but are shown
here to demonstrate consistency with the measurements at
lower redshifts.}\label{ADD}
\end{figure}

\subsection{The $R_{\rm h}=ct$ Universe}
The $R_{\rm h}=ct$ Universe also has only one free parameter, $H_{0}$.
The results for the $R_{\rm h}=ct$ Universe are shown in Figure~1
(dashed line). We see here that the best fit corresponds to $H_{0}=
72.6_{-3.4}^{+3.8}$ km $\rm s^{-1}$ $\rm Mpc^{-1}$ ($68\%$ confidence interval).
With $38-1=37$ degrees of freedom, the reduced $\chi^{2}$ in
$R_{\rm h}=ct$ is $\chi_{\rm dof}^{2}=29.07/37=0.786$.

To facilitate a direct comparison between $\Lambda$CDM and $R_{\rm h}=ct$,
we show in Figure~4 the 38 \emph{Chandra}/SZE cluster distance measurements,
together with the best-fit theoretical curve in the $R_{\rm h}=ct$ Universe
(corresponding to $H_{0}=72.6_{-3.4}^{+3.8}$ km $\rm s^{-1}$ $\rm Mpc^{-1}$),
and similarly for $\Lambda$CDM (with $H_{0}=73.9_{-9.5}^{+10.6}$ km $\rm
s^{-1}$ $\rm Mpc^{-1}$ and $\Omega_{\rm m}=0.50$). In this figure, we
also show the angular diameter distances of nearby clusters from Mason
et al. (2001). These are not included in the fits, but demonstrate
that the best-fit curves are in agreement with low-redshift
X-ray/SZE measurements. Strictly based on their
$\chi^{2}_{\rm dof}$ values, $\Lambda$CDM and $R_{\rm  h}=ct$ appear to fit
the data comparably well. However, because these models formulate their
observables (such at the angular diameter distance in Equations~2 and 3)
differently, and because they do not have the same number of free parameters,
a comparison of the likelihoods for either being closer to the `true' model
must be based on model selection tools. As we shall see, the results of our
analysis in this paper tend to favour $R_{\rm h}=ct$ over $\Lambda$CDM,
which we now demonstrate quantitatively.

\subsection{Model Selection Tools}

The use of model selection tools in one-on-one cosmological model comparisons
has already been discussed extensively in the literature (see, e.g., Liddle
et al. 2006, Liddle 2007). More recently, the successful application
of model selection tools, such as the Akaike Information Criterion (AIC), the Kullback
Information Criterion (KIC), the Bayes Information Criterion (BIC), has been
reported by Shi et al. (2012) and Melia \& Maier (2013). For each fitted model,
the AIC is given by
\begin{equation}
\begin{split}
\rm AIC=\chi^{2}+2\eta,
\end{split}
\end{equation}
where $\eta$ is the number of free parameters. If there are two models
for the data, $\mathcal{M}_1$ and $\mathcal{M}_2$, and they have been
separately fitted, the one with the least resulting AIC is assessed as the
one more likely to be ``true.''  If ${\rm AIC}_\alpha$ is associated with
model~$\mathcal{M}_\alpha$, the unnormalized confidence that
$\mathcal{M}_\alpha$~is true is the ``Akaike weight'' $\exp(-{\rm AIC}_\alpha/2)$,
with likelihood
\begin{equation}
\label{eq:lastAIC}
{\cal L}(\mathcal{M}_\alpha)=
\frac{\exp(-{\rm AIC}_\alpha/2)}
{\exp(-{\rm AIC}_1/2)+\exp(-{\rm AIC}_2/2)}
\end{equation}
of being closer to the correct model. Thus, the difference ${\rm AIC}_2
\nobreak-{\rm AIC}_1$ determines the extent to which $\mathcal{M}_1$ is favoured
over~$\mathcal{M}_2$.

KIC takes into account the fact that the probability density functions of the
various competing models may not be symmetric. The unbiased estimator for the
symmetrized version (Cavanaugh 1999, 2004) is given by
\begin{equation}
\begin{split}
\rm KIC=\chi^{2}+3\eta\;.
\end{split}
\end{equation}
It is very similar to the AIC, but strengthens the dependence
on the number of free parameters (from $2\eta$ to $3\eta$). The
strength of the evidence in KIC favouring
one model over another is similar to that for AIC; the
likelihood is calculated using the same Equation~(7), though
with ${\rm AIC}_\alpha$ replaced with ${\rm KIC}_\alpha$.
The Bayes Information Criterion
(BIC) is an asymptotic ($N\to\infty$) approximation to the outcome of a
conventional Bayesian inference procedure for deciding between models
(Schwarz 1978), defined by
\begin{equation}
\label{eq:BIC}
  {\rm BIC} = \chi^2 + (\ln N)\,\eta,
\end{equation}
where $N$ is the number of data points. It suppresses overfitting very strongly
if $N$~is large, and has now been used to compare several popular models
against $\Lambda$CDM (see, e.g., Shi et~al.\ 2012).

With the optimized fits we have reported in this paper, our analysis of the
cluster angular-diameter distances shows that the AIC does not favour either
$R_{\rm h}=ct$ or the concordance model when we assume prior values for all
of its parameters (except for the Hubble constant). The calculated AIC likelihoods
in this case are $\approx 51.7\%$ for the former, versus $\approx 48.3\%$ for
the latter. However, if we relax some of the priors, and allow both $\Omega_{\rm m}$
and $H_0$ to be optimized in $\Lambda$CDM, then $R_{\rm h}=ct$ is favoured over
the standard model with a likelihood of $\approx 72.5\%$ versus $27.5\%$ using AIC,
$\approx 81.3\%$ versus $\approx 18.7\%$ using KIC, and $\approx 85.7\%$ versus
$\approx 14.3\%$ using BIC. The ratios would be much greater for variants
of $\Lambda$CDM that contain more free parameters than the basic $\Lambda$CDM
model. Since we fixed $\Omega_{\rm m}$ in the $w$CDM model,
there are 2 free parameters when we are fitting the ADD data in $w$CDM.
We find that $R_{\rm h}=ct$ is favoured over $w$CDM by a
likelihood of $\approx 72.8\%-85.9\%$ versus $27.2\%-14.1\%$ using these three model
selection criteria.

\section{Discussion and Conclusions}
In this paper, we have used the cluster angular-diameter size versus redshift data
to compare the predictions of several cosmological models. We have individually
optimized the parameters in each case by minimizing the $\chi^{2}$ statistic. We 
have found that the current cluster ADD constraints are not very restrictive
for $w$CDM, though its optimized parameter values appear to be consistent at
the $1\sigma$ level with those of the concordance model. 

A comparison of the $\chi^2_{\rm dof}$ for the $R_{\rm h}=ct$ Universe and
$\Lambda$CDM shows that the cluster angular-diameter distance data favour
the former over the latter though, on the basis of these $\chi^2_{\rm dof}$
values, one would conclude that both provide good fits to the observations.
The $R_{\rm h}=ct$ Universe fits the data with $\chi^2_{\rm dof}=0.786$ for
a Hubble constant $H_{0}=72.6_{-3.4}^{+3.8}$ km $\rm s^{-1}$ $\rm Mpc^{-1}$,
and $H_{0}$ is the sole parameter in this model. By comparison, the optimal
flat $\Lambda$CDM model, which has two free parameters (including
$\Omega_{\rm m}=0.50$ and $H_{0}=73.9_{-9.5}^{+10.6}$ km $\rm s^{-1}$ $\rm
Mpc^{-1}$), fits the angular-size data with a reduced $\chi^2_{\rm dof}=0.806$.
However, statistical tools, such as the Akaike, Kullback and Bayes Information
Criteria, tend to favour the $R_{\rm h}=ct$ Universe. Since $\Lambda$CDM
(with assumed prior values for $k$ and $w$) has one more free parameter than
$R_{\rm h}=ct$, the latter is preferred over $\Lambda$CDM with a likelihood
of $\approx 72.5\%$ versus $\approx 27.5\%$ using AIC, $\approx 81.3\%$
versus $\approx 18.7\%$ using KIC, and $\approx 85.7\%$ versus $\approx 14.3\%$
using BIC. The ratios would be greater for the other variants of $\Lambda$CDM
because they each have more free parameters than the basic $\Lambda$CDM model.
Since we fixed $\Omega_{\rm m}$ in the $w$CDM model, 
there are 2 free parameters when we are fitting the ADD data in $w$CDM.
We find that $R_{\rm h}=ct$ is favoured over $w$CDM by a
likelihood of $\approx 72.8\%-85.9\%$ versus $27.2\%-14.1\%$ using these three model
selection criteria.

But as we have found in other one-on-one comparisons, fits
to the data using $\Lambda$CDM often come very close to those of
$R_{\rm h}=ct$, which lends some support to our inference that the
standard model functions as an empirical approximation to the latter.
Though it lacks the essential ingredient in $R_{\rm h}=ct$---the equation
of state $p=-\rho/3$---it nonetheless has enough free parameters one
can optimize to produce a comparable fit to the observations. There is
no better example of this than the comparison of curves in Figure~4.
The optimized $\Lambda$CDM prediction tracks that of the $R_{\rm h}=ct$
Universe almost identically. For this particular data set, the difference
in outcome using model selection tools is therefore almost entirely due
to the fact that $R_{\rm h}=ct$ has fewer parameters than the standard model.

With this result, we can now address the second goal of our paper---to provide
some evidence in support of, or against, the strong evolution in galaxy
size implied by earlier studies assuming $\Lambda$CDM. The analysis of the
average linear size of galaxies with the same luminosity, though over a
range of redshifts (see, e.g., McIntosh et al. 2005; Barden et al. 2005;
Trujillo et al. 2006), had indicated that the standard model is consistent
with these data only if these galaxies were $\sim 6$ times smaller at $z=3.2$
than at $z=0$. Lop\'ez-Corredoira (2010) reconsidered this question, pointing
out that current ideas invoked to explain this effect probably cannot adequately
account for all of the deficit of large objects at high redshifts. For
example, the main argument in favour of the evolution in size for a fixed
luminosity is that younger galaxies tend to be brighter, and we should
expect to see younger galaxies at high redshift. Therefore, for a given
luminosity corresponding to some radius at present, galaxies in the past
could have produced the same luminosity with a smaller radius. Lop\'ez-Corredoira
estimates that this effect can contribute a factor $\sim 2-3$ difference in size,
depending on whether one is looking at elliptical or disk galaxies. Other
factors contributing to the evolution in size include mergers, and
observational selection effects, such as extinction.

It appears that none of these explanations, on their own, can account for
the required overall growth. However, our use of galaxy clusters in this paper has
shown that the cluster ADD data are quite consistent with both $\Lambda$CDM
and $R_{\rm h}=ct$. Thus, the fact that both of these models provide an
excellent explanation for the ADD measurements using clusters casts
some doubt on the suggestion that the inferred galaxy-size evolution
is due to the adoption of an incorrect cosmology, rather than effects
such as those discussed above. If it turns out that the combination
of growth factors together cannot account for the factor six difference
in size between $z=3.2$ and $z=0$, we would conclude that some other
astrophysical reasons are responsible, rather than the adoption of
an incorrect expansion scenario.

In conclusion, even though the current galaxy-cluster angular size
versus redshift data are still somewhat limited by relatively large
uncertainties, we have demonstrated that they can nonetheless already
be used to carry out meaningful one-on-one comparisons between
competing cosmologies. The ADD measurements for these structures
appear to be consistent with both $\Lambda$CDM and the $R_{\rm h}=ct$
Universe, suggesting that the inconsistent results obtained with
similar work using individual galaxy sizes are probably not an
indication of gross deficiencies with the assumed cosmological model.

\section*{Acknowledgments}
We acknowledge Yun Chen for his kind assistance.
This work is partially supported by the National Basic Research Program (``973" Program)
of China (Grants 2014CB845800 and 2013CB834900), the National Natural Science Foundation
of China (grants Nos. 11322328 and 11373068), the One-Hundred-Talents Program,
the Youth Innovation Promotion Association, and the Strategic Priority Research Program
``The Emergence of Cosmological Structures" (Grant No. XDB09000000) of
the Chinese Academy of Sciences, and the Natural Science Foundation of Jiangsu Province.
F.M. is grateful to Amherst College for its support through
a John Woodruff Simpson Lectureship, and to Purple Mountain Observatory in
Nanjing, China, for its hospitality while this work was being carried out.
This work was partially supported by grant 2012T1J0011 from The Chinese Academy
of Sciences Visiting Professorships for Senior International Scientists, and
grant GDJ20120491013 from the Chinese State Administration of Foreign Experts
Affairs.

\end{document}